\documentstyle[prl,twocolumn,aps]{revtex}

\begin{document}
\draft
\preprint{}
\title{
Magneto-Optics of type-II superconductors}
\author{E. Choi, H.-T. S. Lihn, and H. D. Drew}
\address{
Department of Physics, University of Maryland,
College Park, Maryland 20742-4111
}
\author{T. C. Hsu}
\address{
AECL Research, Chalk River Laboratories,
Chalk River, Ontario, Canada, K0J 1J0.
}
\date{\today}
\maketitle
\begin{abstract}
The magneto-optical activity of superconducting
${\rm YBa}_{2}{\rm Cu}_{3}{\rm O}_{7}$
observed by Karrai {\it et al.}
is not present in many commonly
employed models of vortex dynamics.
Here we propose a simple, unifying picture for the frequency
dependent magneto-optic response of type-II superconductors
at low temperatures.  We bring together Kohn's theorem,
vortex core excitations, and vortex pinning
and damping into a
single expression for the conductivity tensor.
The theory describes magneto-optical activity
observed in
infrared transmission measurements of thin films of
${\rm YBa}_{2}{\rm Cu}_{3}{\rm O}_{7}$.
\end{abstract}
\pacs{PACS numbers: 74.25.Nf, 74.60.Ge, 74.25.Gz, 76.40.+b}

\narrowtext
For applications of superconductivity vortex
motion and pinning is of great importance. Many models for
vortex dynamics
have been developed such as those of
Bardeen and Stephen \cite{BARDEEN},
Nozieres and Vinen \cite{NOZIERES},
and Clem and Coffey \cite{CLEM}.
One way of studying vortex dynamics is to go to
frequencies that are large compared with
the characteristic frequencies associated with
pinning and damping
to observe the free inertial response of the vortices.
This regime is not well studied and moreover
a microscopic approach is needed. We note that, for
high temperature superconductors, estimates \cite{YEH} of
the frequency range over which one expects electromagnetic
absorption ($10^{10} - 10^{14}~{\rm s}^{-1})$ overlaps with
the microscopic energy scale associated with
quantized
quasiparticle states in vortex cores \cite{CAROLI}.
Because of the short coherence length in high temperature
superconductors, the energy scale of quasi-particle
states localized at the vortex core is large.

These considerations
have prompted a re-examination
of vortex core states.
Karrai {\it et al.} examined the
magnetic field and frequency dependent infrared transmission
coefficient of thin films of ${\rm YBa}_{2}{\rm Cu}_{3}{\rm O}_{7}$
and found evidence for dipole transitions
in vortex cores \cite{KARRAI1}.
In addition, magneto-optical
activity of the superconducting state was observed
in these experiments \cite{KARRAI2}. For frequencies
above the vortex resonance this chiral response
is consistent with the cyclotron resonance of the mixed state.
According to Kohn's theorem \cite{KOHN},
for an isotropic, homogeneous
electron system in a uniform
applied magnetic field H, the {\it only} excitation produced
by a uniform electrodynamic field is the cyclotron
resonance at the frequency $\omega_{c} = eH/mc$,
where $m$ is the bare band mass.
Thus, for an ideal, pure superconductor,
cyclotron resonance is expected.
{}From these considerations, one of the
requirements of a theory of vortex dynamics is that
it be consistent with Kohn's theorem. Commonly employed
theories \cite{BARDEEN,CLEM,YEH}
fail this test since they are non-chiral.

The matrix elements and selection rules for
the dipole transitions in
vortex cores have been studied by Jank\'o and Shore \cite{JANKO}
and Zhu, Zhang and Drew \cite{ZHU}.
Kopnin \cite{KOPNINA} proposed
a peak in the electromagnetic absorption of pure
superconductors due to these
transitions.
However
Hsu \cite{HSU1,HSU2} showed that these intravortex transitions
are not excited by long wavelength probes in the clean limit.
Instead the collective cyclotron motion of the center of
mass is excited.
The ($q=0$) conductivity is
$\sigma_{xx} = (ne^{2}/m)(\delta(\omega - \omega_c)
+ \delta(\omega + \omega_{c}))/2$.
The cyclotron resonance exhausts the sum rule
$\int_{0}^{\infty} d\omega {\rm Re}~\sigma_{xx} =
(\pi/2)ne^{2}/m$ so that there is no spectral weight at
the bare dipole transition
frequency in the vortex core,
$\Delta^{2}/E_{F} \approx \omega_{c}(H_{c2}/H)$, which is
distinct from $\omega_c$.
Therefore, this theory is consistent with the
Kohn's theorem.

According to Hsu\cite{HSU1,HSU2} pinning activates the
otherwise invisible vortex core excitation.
Pinning also produces a zero frequency delta function
in ${\rm Re}~\sigma_{xx}$.
These features
result from breaking translation invariance, which
invalidates Kohn's theorem.
Consistent with this picture
is the observation \cite{CHOI}
that films of ${\rm YBa}_{2}{\rm Cu}_{3}{\rm O}_{7}$
grown on ${\rm LaAlO}_{3}$
substrates have a much smaller signal
at the vortex resonance than
those grown on Si (with a YSZ buffer). Si/YSZ
is a poorer lattice match, and
more conducive to defect formation.

Our conductivity is based on
identifying coherent excited states of a vortex core
with translation of the core\cite{HSU2}. With this identification
we relate certain low-energy excitations with the
velocity ${\bf v}_{L}$ of the vortex core.
The calculation of the $q=0$ conductivity is based on three equations
derived for the clean limit

\begin{equation}
{\bf {\bar J}} = ne({\bf v}_{S}
+ \Phi({\bf v}_{L}-{\bf v}_{S}))~,
\label{XCURRENT}
\end{equation}

\begin{equation}
{{e{\cal E}}\over m}
={\dot{\bf v}}_{S}
- N_{v}(h/2e){\bf v}_{L}\times {\hat z}
={\dot{\bf v}}_{S}
- \omega_{c}{\bf v}_{L}\times {\hat z}~,
\label{ELECTRIC}
\end{equation}

\begin{equation}
{\dot{\bf v}}_{L} =
{\dot{\bf v}}_{S}
- (1-\Phi)\Omega_{0}({\bf v}_{L}-{\bf v}_{S})\times{\hat z}
- {{\bf v}_{L}\over\tau_{v}} - \alpha^{2}{\bf r}~.
\label{ALTEOM}
\end{equation}
${\bf {\bar J}}$ is the
spatially averaged
current density, n is the carrier density, ${\bf v}_{S}$ is the
uniform background superfluid velocity,
$\hbar\Omega_{0} \equiv \Delta^{2}/E_{F} \approx \hbar^{2}/2m\xi^{2}$
is the spacing between quasiparticle levels in the core \cite{BROAD},
$\Delta$ is the bulk gap energy, $E_{F}$ is
the Fermi energy, $\xi$ is the coherence length,
$\Phi \equiv \omega_{c}/\Omega_{0} \approx H/H_{c2}$ is the magnetic
field, $\omega_{c}$ is the cyclotron frequency, ${\cal E}$
is the spatially averaged electric field,
$N_{v}$ is the areal density of
vortices, $\tau_{v}$ is a vortex damping rate,
and $\alpha$ is a pinning frequency.

Eq.~(\ref{XCURRENT}) is obtained by evaluating the
expectation value of the current operator
for a vortex core of velocity ${\bf v}_{L}$ and then
applying Galilean invariance. This ensures that
when ${\bf v}_{L} = {\bf v}_{S}$ the current
is simply that of all carriers moving
at ${\bf v}_{S}$. The ${\bf v}_{L} - {\bf v}_{S}$
part of the current, not present in conventional
theories, comes from coherent excitations of the core \cite{CURRENT_FOOTNOTE}.
In our calculation of the current
due to vortex core excitations
a backflow is not included
when ${\bf v}_{L} \neq {\bf v}_{S}$
since ${\bf\nabla}\cdot{\bf J}({\bf r}) = -d\rho/dt \neq 0$
for the a.c. case.
This differs from the usual
low frequency, quasi-static, vortex motion calculation.
In our calculation we have focused on the
polarizability of the core which couples to an applied
electric field. We have neglected modifications due
to screening.

Eq.~(\ref{ELECTRIC}) says that the total
electric field is the sum of a gauge term from the uniform acceleration
of the superfluid background plus the average Josephson
electric field due to transverse motion of the
vortices. The sign of the latter term
corresponds to a magnetic field
pointing in the $+{\hat z}$ direction
and positive charge carriers.

Eq.~(\ref{ALTEOM}) is an equation of motion for a vortex
core. It was derived in Ref. \cite{HSU2} by looking at
the microscopic equation of motion for low energy quasiparticles
in the core and applying the gap equation. $\tau_{v}$ is
identified as the quasiparticle relaxation time
in the vortex core.
It is the analogue of
the Drude relaxation time for itinerant
electrons.
Note that the damping
in Eq.~(\ref{ALTEOM}) is inversely proportional
to $\tau_{v}$ whereas in theories based on the
Bardeen-Stephen model\cite{BARDEEN,NOZIERES,CLEM} it is
proportional to the normal state
electron transport lifetime $\tau$.
This difference is due to the
clean limit
($\Omega_{0}\tau_{v} \gg 1$) in our theory
and the effectively dirty limit
($\Omega_{0}\tau \ll 1$) conditions of the
hydrodynamic models.
The Bardeen-Stephen model breaks down when $\tau\rightarrow\infty$.
Nevertheless, as we note below,
the two approaches agree in certain cases.
The harmonic pinning term, $-\alpha^{2}{\bf r}$,
can be derived from
a simple single-particle short-range repulsive potential \cite{HSU2}.
It has an appealing and generally expected
form (for low vortex densities).
Notice that in the limit of small dissipation
and pinning the steady state solution of Eq.~(\ref{ALTEOM}) is
${\bf v}_{L} =  {\bf v}_{S}$ as required by Galilean invariance.

Eq.~(\ref{ALTEOM}) contains a term corresponding to a Magnus
force.  The existence of a Magnus force has been
controversial in type II superconductors \cite{NOZIERES}.
Recently, however,
very general arguments have
been given \cite{AO} for a Magnus force even
in the presence of pinning and viscous drag.  A comparison\cite{HSU2} of
Eq.~(\ref{ALTEOM}) to the Magnus force
gives an effective inertial
``mass" of the core
$M_{v} = hn/2\Omega_{0} \approx n\xi^{2}m$
per unit length in the low vortex density limit.
This mass, determined microscopically
from the core energy level spacing, is not equal to
the Suhl value \cite{SUHL,DUAN} for the vortex mass.
Vortex dynamics studies
by Kopnin in
superconductors \cite{KOPNINA} and superfluid
$^{3}{\rm He}$ \cite{KOPNINB}
and by Baym and Chandler \cite{BAYM} in
superfluid $^{4}{\rm He}$ find a vortex mass similar to ours.
The different mass obtained by Suhl appears to be a consequence of the
neglect of quantized core excitations in the Landau-Ginzburg theory.
The electrodynamic response we are considering
corresponds to a polarization of the
vortex core and not the motion of a current pattern
across an otherwise uniform charge density as discussed by Suhl.

The factor $(1 - \Phi)$ in Eq.~(\ref{ALTEOM})
was not derived microscopically but included in order
that ${\rm Re}~\sigma_{xx}(\omega) > 0$ \cite{HSU2}.
This is ensured because Eqs.~(\ref{ELECTRIC}) and (\ref{ALTEOM})
lead to
\begin{eqnarray}
{n\over 2}m&&{d\over{dt}}
\left[\Phi
(\mid{\bf v}_{L}\mid^{2} + \alpha^{2}\mid{\bf r}\mid^{2})
+
(1-\Phi)\mid{\bf v}_{S}\mid^{2}
\right]\nonumber\\
&&=
{\bf {\bar J}}\cdot{\cal E}
-
{{nm\Phi}
\over
{\tau}
}
\mid{\bf v}_{L}\mid^{2}~.
\label{CONSERVATION}
\end{eqnarray}
Taking the time average over one period shows that
$\langle {\bf {\bar J}}\cdot{\cal E}\rangle >0$.
We note that other plausible generalizations of conventional
vortex dynamics theories to
include the vortex mass
generally do not satisfy this condition.
A decreasing $\Omega_{0}$
with magnetic field is expected on general grounds.
It can come from a renormalization of the Magnus force
due to the presence of nearby vortices.
Also the magnetic field
increases as the vortices overlap and will affect
the dynamics. These effects
were ignored in our single vortex calculation.

The three equations in four unknowns (${\bf {\bar J}}$, ${\cal E}$,
${\bf v}_{L}$, ${\bf v}_{S}$)
may be used to eliminate
${\bf v}_{L}$ and ${\bf v}_{S}$
leaving
${\bf {\bar J}} = {\bf\sigma}{\cal E}$
with
\begin{equation}
\sigma_{\pm}={{ine^{2}}\over{m\omega}}
\left[
{{
\omega
(\omega\mp{\bf{\bar{\rm\Omega}}}_{0})
+ (1 - \Phi)
(i\omega/\tau_{v}-\alpha^{2})
}\over
{
(\omega\mp{\bf{\bar{\rm\Omega}}}_{0})
(\omega\pm{\omega}_{c})
+i\omega/\tau_{v}-\alpha^{2}
}}
\right]~,
\label{CONDUCTIVITY}
\end{equation}
where $\sigma_{\pm} = \sigma_{xx} \pm i\sigma_{xy}$
and
${\bf{\bar{\rm \Omega}}}_{0} = (1-\Phi)\Omega$.
The ${\pm}$ refers to cyclotron resonance active $(-)$
or inactive $(+)$ modes of circularly polarized light.
This conductivity satisfies a number of
important limits. First, when $B=0$
the London conductivity is recovered.
For zero pinning and zero dissipation
the conductivity contains simply
the cyclotron resonance as given above.
In the limit of no pinning,
zero frequency, and low vortex density,
$\rho_{xx} = (m/ne^{2}\tau_{v})\Phi = \rho_{xx}^{n}\Phi$
($\rho_{xx}^{n}$ is the normal state resistance)
since $\tau_{v}$ is identified with $\tau$.
Therefore we obtain the same expression in
the clean limit as was found in the hydrodynamic
theories \cite{BARDEEN,NOZIERES}.
In this limit we also obtain
$\rho_{xy} = B/nec$ as expected from Galilean invariance.
The sum rule is satisfied. For example pinning produces
a delta function,
\begin{equation}
{\rm Re}\ \sigma_{xx}(\omega)
= {{ne^{2}}\over{m}}
\left[{{
(1 - \Phi)\alpha^{2}
}\over
{\alpha^{2} + \omega_{c}{\bf{\bar{\rm\Omega}}}_{0}}
}
\right]
\pi\delta(\omega)~.
\label{DELTA_FUNCTION}
\end{equation}
This spectral weight is removed
from the cyclotron resonance. The cyclotron
resonance is shifted to
$\omega_{c} + \alpha^{2}/{\bf{\bar{\rm\Omega}}}_{0}$
to produce a hybrid cyclotron-pinning resonance.
Spectral weight is
also transfered between the cyclotron resonance and the vortex core
resonance.  In Fig.~\ref{cond_fig} we plot the conductivity
and show how it is affected by pinning.
For values of the pinning frequency that are consistent with
experiments our
theory predicts a very small optical response at
${\bf{\bar{\rm\Omega}}}_{0}$
in the cyclotron inactive mode.

If we calculate the thin film surface impedance, ${\cal Z}$, in the
low frequency limit
($\omega \ll \alpha ,\Omega_{0}$) and
the limit where pinning dominates the magnus force
($\alpha^{2} \gg \omega\Omega_{0}$)
we obtain the Gittleman-Rosenblum\cite{GITTLEMAN},
Coffey-Clem\cite{COFFEY} result for pinned
vortices. Alternatively we could derive this
by dropping the ${\bf{\dot v}}_{L}$,
${\bf{\dot v}}_{S}$, and ${\bf v}_{L}\times{\hat z}$
terms in Eq. (\ref{ALTEOM}). Dropping the ${\bf{\dot v}}_{L}$
term is equivalent to setting the mass to zero.
Using ${\cal Z} = 1/\sigma_{\pm}(\omega)t$, ($t$ is the film
thickness), and defining
a drag coefficient $\eta = M_{v}/\tau_{v}$,
and a pinning force constant $\kappa=M_{v}\alpha^{2}$,
\begin{equation}
{\cal Z} \approx
{{
4\pi\lambda^{2}\omega}
\over
{tc^{2}i(1-\Phi)}
}
+
{{
B\Phi_{0}}
\over
{tc^{2}(\eta + i\kappa/\omega)}}~.
\label{IMPEDANCE}
\end{equation}
The first term is
the London reactance term. It differs from the Coffey-Clem result
by the $1-\Phi$ factor. This reduction of the effective
superfluid density  by $1 - (H/H_{c2})$ is not unexpected
nor contradicted by experiment.
Eq. (\ref{IMPEDANCE}) was obtained under the
clean limit assumptions
of our theory. Remarkably it has the same form as the
dirty limit case. Therefore if the Magnus force were
present in the dirty limit, as argued by
Ao and Thouless \cite{AO}, then the correct conductivity
function in that limit might also resemble Eq. (\ref{CONDUCTIVITY}).

We have fit measurements of the infrared transmission of thin
films of ${\rm YBa}_{2}{\rm Cu}_{3}{\rm O}_{7}$
with this theoretical conductivity.
The experimental technique is described
in Refs. \cite{KARRAI1,KARRAI2,CHOI}.
The transmission coefficient is
$T^{\pm} = 4N/|N+1+(4\pi/c)\sigma_{\pm}t|^{2}$.
$N$ is the refractive index
of the substrate.
In Fig.~\ref{tplusminus} we plot the ratio of the
transmission coefficients $T^{+}$ and $T^{-}$.
This ratio mostly cancels out non-chiral components of the
transmission.
The figure shows a definite optical activity in the sample
which at high frequencies
is well described by a lossless free
electron conductivity \cite{KARRAI2,CHOI}.
At frequencies below $\sim50~{\rm cm}^{-1}$, however,
the chiral response cannot be well represented even
if damping is included in the free electron conductivity
as can be seen
in Fig.~\ref{tplusminus}. In particular
there is a steep drop in $T^{+}/T^{-}$ at low
frequencies which corresponds to the onset of the hybrid
cyclotron-pinning resonance.
This feature is
well fit by our theoretical conductivity.
The free parameters in our fits are
$\tau_{v}$, $\alpha$, and $\Omega_{0}$.
We note that the conventional conductivity function
\cite{BARDEEN,NOZIERES,CLEM}, even when
generalized to finite vortex mass,
is non-chiral and would predict simply $T^{+}/T^{-} = 1$.
That $T^{+}/T^{-}$ falls below unity at
$30~{\rm cm}^{-1}$ is particularly noteworthy as it implies
a sign reversal of the a.c. Hall effect at low frequencies.

Another observation from the data
displayed in Fig.~\ref{tplusminus} is that
the chiral signal associated with
the vortex core resonance is very small compared
with the cyclotron-pinning resonance.
Our theory also predicts a very small chiral
signal but does not agree in detail with the
measured signal.
The measured resonance is larger than the theory predicts.
Moreover, there are substantial non-chiral features observed
in the transmission spectrum around $65~{\rm cm}^{-1}$.
These features are brought out in
unpolarized transmission as can be seen in
Fig.~\ref{linear_polarized}. The peak around $65~{\rm cm}^{-1}$
is substantially larger in the unpolarized spectrum
than in the $T^{+}/T^{-}$ spectrum.
This figure also shows the theoretical unpolarized transmission.
The corresponding small broad peak in the calculated curve
is chiral
and has been adjusted to have the maximum
amplitude consistent with the low frequency rise in the transmission.
Our conductivity does not contain
the non-chiral response observed in the experiments. Nevertheless, we
believe \cite{CHOI} that this non-chiral
response and the enhanced chiral response
may be due to vortex core excitations.
The effects of defects beyond harmonic pinning
induce optical transitions which, because of the breakdown of the
cylindrical symmetry, have less restrictive optical selection rules.

The presence or absence of polarization
dependence is controversial. The considerations
of Zhu, Zhang and Drew \cite{ZHU} and Jank\'o and
Shore \cite{JANKO} would predict that the
vortex resonance
is in the cyclotron active mode. The conductivity here leads
to the opposite effect: a weak vortex resonance in the cyclotron resonance
inactive mode.

In conclusion, we have presented a conductivity function
which produces magneto-optical activity at high frequencies and
explains how it is diminished by vortex pinning and damping
at low frequencies.  The theory
predicts that the dipole-allowed vortex resonance is very weak due to strong
screening by
the vortex motion even in the presence of
pinning. Thus, the Kohn theorem is very robust in this system.
The conventional theories of
vortex dynamics are not consistent with Kohn's
theorem and do not contain optical activity.
Therefore our conductivity function provides a better starting point
for studying the interaction of
vortices with impurities and the
excitation of vortex core states.

We thank Qi Li and D.B. Fenner for providing the samples
used in this study.
Useful discussions with S. Anlage, S. Kaplan, K. Karrai, C. Lobb, N.P. Ong,
J. Orenstein, B. Parks, S. Spielman,
and F.-C. Zhang are acknowledged.  This work was supported
in part by the National Science Foundation
under grant No. DMR 9223217.

\begin{figure}
\caption{The frequency dependence of the conductivity function,
Eq.~(5), is
plotted for cyclotron inactive $(+)$ and active $(-)$ modes.
The parameters
are, $\omega_{c} = 10~{\rm cm}^{-1}$,
$\alpha = 20~{\rm cm}^{-1}$,
$\Omega_{0} = 60~{\rm cm}^{-1}$, and
$1/\tau_{v} = 10~{\rm cm}^{-1}$. There is also a delta
function at zero frequency
whose strength is given by Eq.~(6).}
\label{cond_fig}
\end{figure}
\begin{figure}
\caption{The chiral response for the
${\rm YBa}_{2}{\rm Cu}_{3}{\rm O}_{7}/{\rm Si}$
sample at $H = \pm 12~{\rm T}$ and 2.2~K as
determined from measurements with a polarizer with a 0.9~mm quartz
wave plate.  The solid curve is the best fit to our
conductivity function.  The best fit parameters are
$\alpha = 50\pm 2~{\rm cm}^{-1}$,
$1/\tau_{v} = 40\pm 5~{\rm cm}^{-1}$,
when $\Omega_{0}$ was taken as $60~{\rm cm}^{-1}$ and
the cyclotron mass m was taken to be 3.1 electron
masses [7].
The dashed curve is
the best fit to the damped free electron model.  The corresponding
damping paramter is $1/\tau = 56~{\rm cm}^{-1}$.}
\label{tplusminus}
\end{figure}
\begin{figure}
\caption{Transmission ratio at 2.2~K for the
${\rm YBa}_{2}{\rm Cu}_{3}{\rm O}_{7}/{\rm Si}$
sample taken at 12~T and
2.2~K in unpolarized light.  The solid line is the
theoretical result using the
best fit parameters from the analysis of the chiral data in Fig.~2.}
\label{linear_polarized}
\end{figure}

\begin{references}
\bibitem{BROAD} We assume that there is little
	quasiparticle energy level
	dispersion parallel to the vortex lines. This
	is the case when vortices in
	${\rm YBa}_{2}{\rm Cu}_{3}{\rm O}_{7}$
	lie perpendicular to the Cu-O planes).
\end{references}
\end{document}